\documentclass[superscriptaddress,eqsecnum,amsfonts,showpacs]{revtex4}

\usepackage{epsfig}

\newcommand{\be}{\begin{equation}}

\newcommand{\ee}{\end{equation}}

\newcommand{\bea}{\begin{eqnarray}}

\newcommand{\eea}{\end{eqnarray}}

\newcommand{\nn}{\nonumber}

\newcommand{\ep}{i\epsilon}

\newcommand{\om}{\omega}

\begin{document}

\title{Confinement within the use of Minkowski integral representation}

\author{V. \v{S}auli}

\affiliation{Department of Theoretical Physics, NPI Rez near Prague, Czech Academy of Sciences}

\begin{abstract}

We determine  the gluonic spectral function $SU(3)$ Yang-Mills theory
as well as we found fermionic spectral functions in the  strong quenched QED where
 we found new solutions. Our novel technique provides solutions with the usual branch cut for propagators
 while not showing any pole structure at the first Riemann sheet (identical with entire complex  plane)
 of complex  square of momentum.  Implications and further utilizations are briefly adressed for QCD and Standard model
 calculations.

\end{abstract}

\maketitle

\section{Introduction}

Confinement is one of essential features of QCD, and whilst it  is intuitively well understood as impossibility to 
 observe the excitations of quark and gluon fields as a particle  in isolation, a commonly accepted mathematically rigorous
explanation of this phenomena is still missing.

Apart of QCD,  $QED_{2+1}$ as known from many studies \cite{ANW1988,AKO1987,BPR1992,FAD2004,Ma1995,DOMA1992,FRTE2001,HE2005}
 is a simple theory showing confinement. Notably, it serves as  an effective theory in condense matter
  physics and physics of graphene \cite{GSC2007} for many years.
  Here we argue, that for  the coupling is strong enough,  confinement exists in quenched $QED_{3+1}$ as well. The value
 of the coupling where the fermion stop from being free particles  coincides with the critical coupling of chiral symmetry breaking 
 in theory with hard cutoff. 
  
In presented paper we  revise the numerical method based on utilization of  integral representation for
 $QED$ and  $SU(3)$ Yang-Mills theory, more complicated issue of confinement of quarks in entire  QCD theory  is briefly  
 mentioned, but not solved here.

%Reflecting the difficulty of calculation performance, an overwhelming majority of canonical Schwinger-Dyson equations (SDEs)
%\cite{COR1982,AGPA2006,, cornwall2009,BIPA2009,AGBIPA2011,AGBIPA2015,ABFP,ABP2017,AFTP2019,FIHU2020,ABP2017,alk2021}
%studies exist  in the Landau  gauge, while a few studies consider a larger class of linear gauges \cite{ABP2017,alk2021}. To the date

In principle,  theoretical search for a   confined modes in QFT 
 could be  equivalent for proving of  the absence of particle modes
 in the  time ordered product  of associated field: the quark and the gluon propagators, in QCD for instance.
  For this purpose we follow the integral representation methods   \cite{SAJHEP,COR1982,VSJA2003,DEWE1977,SAUFBS,trans,SAB2007,MESA20,sau2,MS2021,sau1,YPNFS2021,sau3,DFPY2022,SAULI}
 developed for purpose of solution of Schwinger-Dyson equations (SDEs).
These methods offer the  solution in the entire Minkowski space and the presence or absence of particle modes
can be very obvious: one can or one does not  see the  on mass shell pole in the propagator
 (unphysical poles that do not appears in S-matrix are allowed, they can emerge in special  choices of gauge and renormalization schemes 
 and we do  not discussed them here. In  models considered here,  the only example is the  massless pole in longitudinal part of the gauge propagator).
  Owing to other gauges a different confinement pictures have emerged in history.
The Gribov-Zwanziger picture of confinement \cite{GR1978},\cite{ZW1989}  
is  established in Coulomb gauge \cite{ERS2007}, while not confirmed in the class of linear gauges (e.g. in Landau gauge), 
where the lattice simulation of ghost  propagator does not show  
 enhancement- the required condition for the Gribov-Zwanziger scenario. Other scenario of  confinement- 
 the vortex condensation \cite{1a,1b,1c,1d,1e,1f} with associated  condensation of magnetic monopoles are 
detected on lattices in special gauges \cite{4},\cite{5},\cite{6}, e.g. in maximal Abelian gauge \cite{7}.
Obviously, different manifestation and realizations of confinement exist in various gauges, while they do not show up in other gauges.

To get the resulting form of propagators in the entire Minkowski space, there are several existing  methods in hand.
In principle, one can apply  Schlessinger method of analytical continuation on the already known solution of
 SDEs obtained by conventional method in the Euclidean space \cite{ZRC2021,BT2019} (for the topic of SDEs
 see \cite{osud,AS2001}). In the study \cite{ZRC2021}  the method confirms 
well known fact that the analytical property are  affected by the truncation of SDEs , e.g. by the analytical properties of  kernels which approximate 
unknown Green's functions that have been truncated out of the  SDEs system.
Actually, there exist well known models, where number of complex conjugated poles 
are supposed to be a part of the spectrum of calculated propagator. 
 Using other methods, the analytical continuation of lattice  propagators
 has been questioned  in  papers \cite{DOS2014,TGUS2019}, showing the answer is still very ambiguous.
 Furthermore, complex poles necessarily generate complex branch points via quantum loops, thus  they can never appear alone
 and they distort usual form of dispersion relation \cite{HPW2022}, making a correct calculations more complex and more difficult.
Besides the theories and models exhibiting confinement clearly,
 the spectrum of quarks does not need to show up confinement in some effective 
models. For instance  quark-meson model studied in \cite{TWSW2018} provides a typical on-shell delta function 
in the spectral function of the quark propagator and  such  model does not explain confinement at all.

To get the analytical continuation in our presented study, 
we convert the  DSEs system ( written in momentum space) into a new system of equations for spectral
 functions as the first step. Our main goal is developing  numeric, 
 which makes the original method  developed by the author earlier in \cite{SAJHEP,VSJA2003,SAUFBS,SAULI}
 actually working and useful in practice  for  strongly coupled theories with confinement. Actually, this second step 
 was lacking for decades and this completion is the main purpose of presented study.

 The paper’s organization is as follows.
In the next Section II we explain how to get spectral function within the simple model: the quenched QED in 3+1 dimensions.
 In the Section III  we revise numeric used in \cite{SAULI} and offer a more precise solution for the gluon DSE there. We also make a short remark
 on the existing spectral solutions for the quark propagator in the Section IV.

\section{Strong Quenched QED in 3+1 dimensions}

 Weakly coupled quenched QED in $3+1$ dimension is certainly a theory without confinement. It contains free moving electrons in its spectrum and 
   for $\alpha$ satisfying inequality $\alpha<\alpha_c$ one gets non-confining spectral solution for the fermion propagator \cite{SAJHEP}. 
   The spectral function has a delta function associated with free
 propagation and continuum part above.  We argue, that such solution  does not exist for super-critical couplings 
 (i.e. for the value  of the coupling such that $\alpha>\alpha_c=\pi/3$), for which 
  case we provide another -confining- spectral  solution for the first time. 
  We avoid the use  of hard cutoff,  introduce the electron mass term explicitly.  There would be cut off dependent interplay between two effects chiral symmetry breaking and confinement. Remind the reader the mass function  obey  Miransky scaling  $M(0)\simeq \Lambda exp(\alpha-\pi/3)^{-1/2}$
, which   persists in more sophisticated approximations \cite{bashir}. 
    Therefore, since our fermion is explicitly massive in presented study,  we  can use several equivalent computational schemes, including dimensional regularization/renormalization
   \cite{THVE1972}, BPHZ scheme \cite{BPHZ1} as well as an L-operation scheme \cite{sauL}, 
   all incorporate mass renormalization as usually.  

Conventional renormalization schemes (dimensional or BPHZ)  together  with simultaneous  use of  the  following spectral representation:
\bea \label{SR}
S(k)=\int_0^{\infty}d a\frac{\not p \rho_v(a)+\rho_s(a)}{p^2-a+\ep} \,
\eea   
 for the fermion propagator $S$ calls for  the  renormalization of  the mass. The reason is that the Euclidean space loop integral turns to be  divergent after changing ordering of integrations.

 Our convention for Minkowski  metric tensor reads: $g_{\mu\nu}=diag(+1,-1,-1,-1)$. Minkowski momenta are not labeled in our 
 notation and we  write index $E$ when we want to specify the  Euclidean momentum ( i.e.  $q^2_E=-q^2$   for the square).

We restrict here to   quenched ladder-rainbow approximation of the fermion DSE, which   in the Landau gauge reads:
 \bea
 S^{-1}&=&\not p-m_o -\Sigma(p) 
 \nn \\
 \Sigma(p)&=&ie^2\int \frac{d^4k}{(2\pi)^4}\gamma^{\mu}S(k)\gamma^{\nu}\frac{P_{\mu\nu}(k-p)}{(k-p)^2+\ep}  \, ,
 \eea
where $e$ is the fermion charge, $P(z)$ is transverse projector. Two  functions or distributions $\rho_{v,s}$ in the Eq. (\ref{SR}) 
are enough to complete two scalar propagator functions $S_{v,s}$ (or $A,B$ alternatively). In our notation they are defined as
\be 
S(p)=S_v(p)\not p+S_s(p)=\frac{1}{\not p A(p^2)-B(p^2)}\, .
\ee

\begin{center}{\bf DSE in dimensional renormalization scheme}\end{center}

Substituting the representation (\ref{SR}) into the DSE, swapping the order of integrations and making the integration over the momentum  one gets
\bea  \label{kuk}
B(p^2)&=&m_0+\frac{3e^2}{(4\pi)^2}\int_0^{\infty}d a \rho_s (a)\int_0^1 dt \left(ln\left[\frac{p^2 t -a+\ep}{\mu_t^2}\right]+C\right)
\nn \\
&=&m(\mu_t)+\frac{3e^2}{(4\pi)^2}\int_0^{\infty}d a \rho_s (a)\int_0^1 dt ln\left[\frac{p^2 t -a+\ep}{\mu_t^2}\right] \, ,
\nn \\
A&=&1 \, ,
\eea
where $\mu_t$  is t’Hooft  renormalization scale of  MS bare  dimensional renormalization scheme, which has been used.

In the next step we will add the zero of the following form
\be
0=B(\zeta)-m(\mu_t)-\frac{3e^2}{(4\pi)^2}\int_0^{\infty}d a \rho_s (a)\int_0^1 dt ln\left[\frac{\zeta t -a+\ep}{\mu_t}\right] 
\ee
to the rhs. of (\ref{kuk}), i.e. we subtract the equation with itself evaluated at the scale $\zeta$. We thus get
\be  \label{proc}
B(p^2)=B(\zeta)+\frac{3e^2}{(4\pi)^2}\int_0^{\infty}d a \rho_s (a)\int_0^1 dt ln\left[\frac{p^2 t -a+\ep}{\zeta t -a+\ep}\right]\, .
\ee
Note also, the function $M=B/A$ is renormalization scheme invariant here as well as in other gauges. 

\begin{center}{\bf  DSE in L- renormalization scheme}\end{center}

In dimensional regularization prescription and the singular pole term   $1/{\epsilon_d}=1/(d-4)$ 
in the constant $C$ is absorbed into the renormalized mass, such that
\be
m(\mu_t)=m_0-\frac{3e^2}{(4\pi)^2}\int_0^1 da \rho_s(a) [1/{\epsilon_d}+\gamma_E +4\pi]
\ee 

Similarly, using the auxiliary Feynman integration one can get the renormalized DSE  equation in the so called $L-$operation scheme \cite{sauL} . The equation is 
 then exactly identical to the second line in (\ref{kuk}) with $\mu^2_t$ replaced by newly introduced scale  $\mu^2_F$ and the renormalized mass
 reads 
\be 
 m(\mu_F)=m_0+\frac{3e^2}{(4\pi)^2}\int_0^1 da \rho_s(a) [ln{\epsilon_z}-ln(\frac{\lambda^2}{\mu^2_F})]
\ee 
 in this scheme. Note that the  scales $\lambda$ as well as $\mu_F$ are arbitrary and the limit $\epsilon_z\rightarrow 0$ is
 understand.  In this scheme the logarithmic divergence presented in the original momentum space integral turns 
 to logarithmic divergent term but now in the Feynman parametric integral.

In order to maintain renormalizability in any known scheme, the mass term  is
 inevitably presented, otherwise  spectral representation could not be used. 
No matter what  the  amount  of dynamically generated mass is, the meaning of dynamical mass generation is always limited
in strong coupling QED.  A pre-defined ordering of limits, which appear in theory,  offers a naive definition of the chiral limit in case of vanishing integral 
\be \label{less}
\int da \rho_s(a)=0 \, .
\ee

Since the quenched QED in 3+1D is not an asymptotically free theory, 
it  splits to different effective models   according to what  regularization method is used.
Massive theory used and renormalized  herein can be hardly compared to cutoff theory considered for instance in  \cite{FUKU1976}. 
Nevertheless, this is simplicity of quenched QED and appearance of confinement through the absence of free propagating modes 
of Lagrangian fermion field, which deserves our attention.

  \begin{center}{\bf The method of solution}\end{center}

For the spacelike value of $p^2 $ variable and for  the negative $\zeta$ we can  drop out Feynman $\ep$ and  solve the equation in the spacelike domain of Minkowski space.
 Note plainly, that  for this purpose one should know the spectral function $\rho$ in advance. 
Therefore, in order  to determine the function  $\rho$ it is helpful to consider the Eq. (\ref{kuk}) at the timelike scale, where  the running mass $B$ is a complex valued function. 
The analytical continuation of the DSE (\ref{proc}) for $p^2>0$ can be written as
\bea  \label{goldfinal}
\Re B(p^2)&=& \Re B(\zeta)+\frac{3e^2}{(4\pi)^2}\int_0^{\infty}d a \rho_s (a)\int_0^1 dt ln\left|\frac{p^2 t -a}{\zeta t -a}\right|\, ,
\nn \\
\Im B(p^2)&=&-\frac{3e^2}{16\pi}\int_0^{\infty}d a \rho_s (a)\left[(1-\frac{a}{p^2})\theta(p^2-a)-(1-\frac{a}{\zeta})\theta(\zeta-a)\right] \, ,
\eea
where $\theta(x) $ is  the usual Heaviside step function, which reads  $\theta(x)=1$ for $x>0$ and $\theta(x)$ is zero otherwise. 

Keeping the equation for the dynamical mass function $B$ in hand, one can reconstruct the propagator $S$. Comparing the real and the imaginary parts of the Eq. 
(\ref{proc}) and the Eq.(\ref{SR}) one can write the following  complementary equation:
\be \label{zdar}
\rho_s(s)=\frac{-1}{\pi}\frac{\Im B(s) R_D(s)+\Re B(s) I_D(s)}{R_D^2(s)+I_D^2(s)}    
\ee
where $s=p^2>0$ in our metric convention, and where the functions  $R_D$ and $I_D$  stand for  the square of the real and the imaginary part of the function $sA^2(s)-B^2(s)$, i.e. 
\bea
R_D(s)&=&s[\Re A(s)]^2-s (\Im A(s))^2-[\Re B(s)]^2+(\Im B(s))^2 \, ,
\nn \\
I_D(s)&=&2 s \Re A(s)\Im A(s) +2 \Re B(s) \Im B(s) \, ,
\eea
  where we keep  $A$ non constant  for general purpose (note, that the Eq. $A=1$ is valid in the approximation employed here).

To get the solution, we start with some initial guess for the constant $\Im B(\zeta) $
as well as for  the trial spectral function   $\rho_s(s)$ and   substitute it  into the Eq. (\ref{goldfinal}). 
Then three equations (\ref{goldfinal}) and (\ref{zdar}) have been solved numerically by the method of iterations.

In this way, after relaxing the  iteration process,  irrespective of the numerical accuracy the obtained function $\rho_s$ is still not what we are looking for. 
The system is ill constrained by our initial  guess for the complex phase  $\phi=tan^{-1} \Im B(\zeta)/\Re B(\zeta)$.
 Obviously the system  of equations has a rich structure of solutions among them we need to find the correct one.
 To get rid of the problem we fix  $\Re B(\zeta) $ and repeat iteration procedure described above for a new value $\Im B(\zeta)$ and look at the quality of equality:
\be \label{equiv}
L(s)=R(s)=\frac{1}{4} \Re Tr S(s)=\Re S_s(s)     
\ee
where
\bea \label{condition}
L(s)&=&P.  \int_0^{\infty}d a\frac{\rho_s(a)}{p^2-a}
\nn \\
R(s)&=&\frac{-\Im B(s) I_D(s)+\Re B(s) R_D(s)}{R_D^2(s)+I_D^2(s)}         
\eea  
 for a given choice of the phase $\phi(\zeta)$. 
 By minimizing the difference $L-R$  we identify a correct value of the phase    $\phi(\zeta)$ (for which the Eq.  $L-R=0$ is hold).
Note, this very similar two steps method has been successfully applied to gain a solution of  DSE of the quark propagator  very recently \cite{sau1,sau2}.

 Similar could be done for the function $S_v$ (see Appendix), however in our approximation  the function $S_v$ is completely determined by the function $S_s$
  and to check the validity of Eq. (\ref{equiv}) is enough.  
\begin{figure} 
\centerline{\epsfig{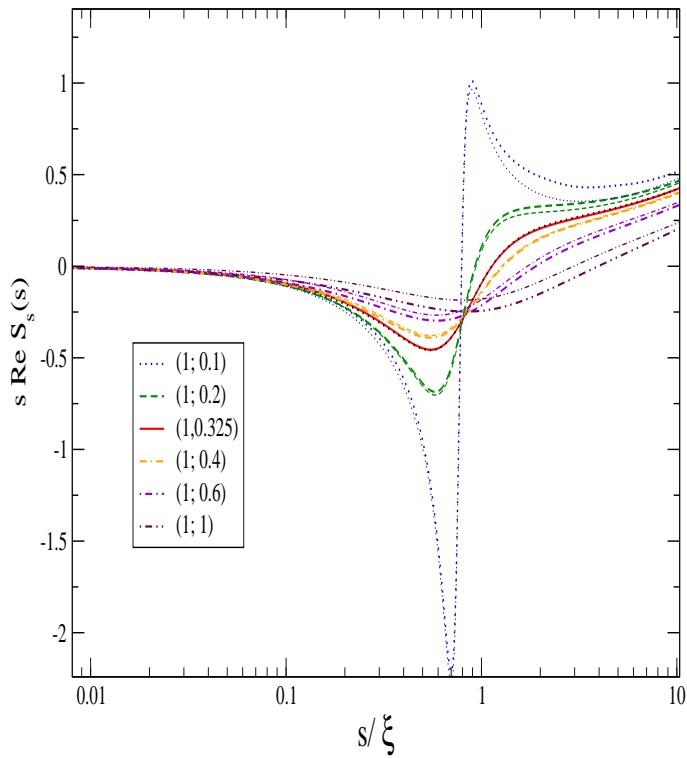}}
\caption[caption]{\label{prvni} $sS(s)$ function constructed from L and R functions. Each type of line corresponds to a particular choice of the phase, 
they are labeled by pair of numbers: real and imaginary part of $B(\zeta)/\zeta$. The exception in labeling is the true DSE solution, i.e. best matching case, 
where  ``L’’ is dotted , while ``R’’ solution is depicted by the solid line.}
\end{figure}
We show the results in the Fig. \ref{prvni}, where the solid line represents result for the propagator corresponding to the ratio $\Im B(0.1)/\Re B(0.1)=0.325 \pm 0.005 $.
The real part of $B(\zeta)$ is  our choice, while  the imaginary part $\Im B(\zeta=1)$ was  the subject of the numerical scan.
In order to visualize our numerical search, we plot two lines  corresponding  to  $L$ and $R$ as defined in by (\ref{condition}). 
More L-line  can be  distinguished R-one, larger deviation from  the equality  (\ref{equiv}) is observed.
 Lines  are labeled  by two numbers representing  values $\Re B(\zeta),\Im B(\zeta)$.

The absence of the real pole in the propagator  is  obvious and instead of the physical threshold the propagator has the zero branch point.
 Confinement is  because of an abrupt generation of  absorptive part of the mass function in the infrared domain of the  timelike momentum.
Propagator shape does not correspond to the decay  of particle mode, but 
is associated with the creation-reabsorption process of confined modes of the fermion quantum field.
It is worthwhile to mention that the generation of zero anomalous threshold is the old conjecture of  Schwinger \cite{SCHWING} made for 
1+1 QED. His idea has emerged before we have accepted QCD as a correct theory of hadrons.

\begin{figure} 
\centerline{\epsfig{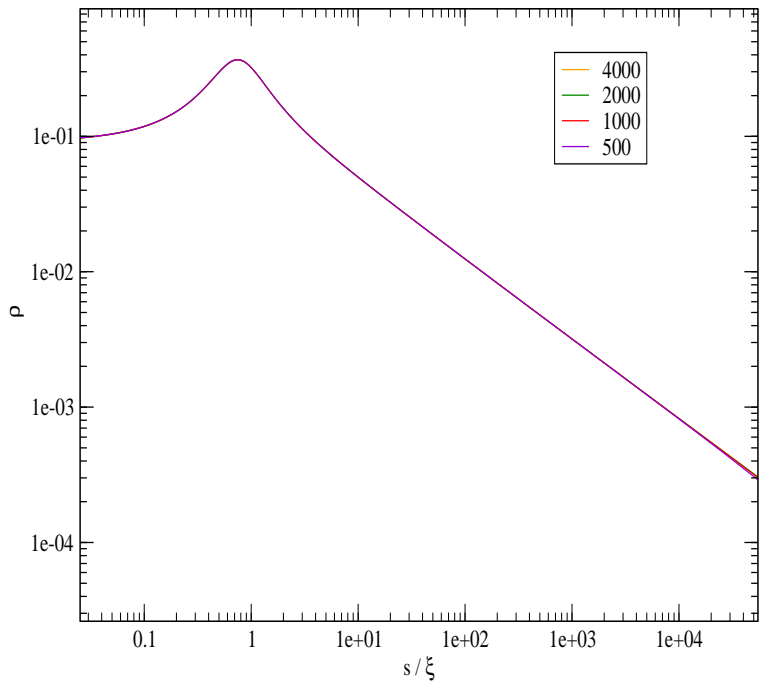}}
\caption[caption]{\label{specfig} Spectral function of strongly interacting fermion in units of renormalized mass $\zeta$.
Exhibition of numerical stability is manifest, note that the labeling of the lines corresponds to the number of integration points,
 no associated cutoff regulator dependence is observed. }
\end{figure}

In non-confining quantum field theory, the  Osterwalder-Schrader axiom of  reflection positivity \cite{OS1975}
 is equivalent to the positive definiteness of the norm in Hilbert space of the corresponding Quantum Field Theory.
 Violation of  reflection positivity  is often regarded as a manifestation of confinement \cite{AS2001}. 
 In the Fig. \ref{specfig} we show the spectral function of the fermion propagator. 
Obviously the property of  reflection positivity is not lost in our case, however the fermion turns to be a short living excitation
according to suggestion made (albeit for the photon) by J. Schwinger half century ago.
Violation of reflection positivity turns out to be a weak criterion for confinement in the model presented here. 

 We do not show the evolution of propagator functions with the coupling. Within decreasing coupling the shape of scalars $S_{v,s}$ gradually rise elbows and the on shell  pole emerges for subcritical value of couplings. Thus for small couplings then one needs also
to determine the pole position and its residuum as  done for theory in even dimensions \cite{SAJHEP},\cite{SAUFBS} as well as in theories  
  with odd number of spacetime dimensions \cite{HOPAWI2006}. 
We also do not go beyond quenched approximation because of the Abelian character of the interaction, but
study of unquenching effects is in principle possible.

As we have already mentioned, we do not get any  truly conformal solution.
 Perhaps, it could  be doable  if  the  superconvergent relation
 $\int \rho(a)=0$ is fulfilled. This condition has never been satisfied and the solution likely does not exist at all.

 \section{Confinement of gluons in pure Yang-Mills theory}

In this section we provide several solutions of SDE for gluon propagator as obtained by analytical
continuation within integral representation method. For our purpose we  will use a simple truncation introduced in
 \cite{BIPA2009}, which was further studied and extended  in \cite{AGPA2006,cornwall2009,AGBIPA2011} in the Euclidean space
as well as the first, not yet perfectly convergent  solution in Minkowski space was attempted in  \cite{SAULI}.
  To remind briefly, our calculation framework is based on nonperturbative generalization of
 the  Background Field method  \cite{PIL1997} and used in Landau gauge for purpose of comparison with 
 lattice data.The goal achieved here is to offer precise solution in the  timelike momentum domain, which  
 is based on the method of subtractions at the timelike scale in an exact manner described 
in the  previous section.

%ZAVER:
 
%To provide further related references on the topic,
 %there is a certain progress \cite{AGBIPA2015,ABP2013,ABFP,ABP2017,AFTP2019,FIHU2020} in developments
 %of DSEs in Yang-Mills theory  in other  gauges then Landau gauge.  Notably, changes within  varied gauge fixing parameter are shown to be dramatic in the %sector of unphysical gluon and ghosst propagators \cite{alk2021}, very likely providing convergent setup for calculation of hadronic 
 %form factors when the quark fields are included.  We briefly mention the progress made in the next section.

For completeness we list all necessary  ingredients of the gluon SDE truncation. The Schwinger mechanism is employed through the simple Anstaz for the unproper  three gluon PT-BFM dressed vertex, which reads 
\be 
d(k)\tilde{\Gamma}_{\nu\alpha\beta}(k,q)d(k+q)=\int d\om \rho(\om)\frac{1}{k^2-\om+\ep}\Gamma_{\nu\alpha\beta}^L
\frac{1}{(k+q)^2-\om+\ep} +d(k)\tilde{\Gamma}_{\nu\alpha\beta}^Td(k+q)
 \, ,
\ee
where $\Gamma_{\nu\alpha\beta}^L$ satisfies tree level WTI and $d$ is scalar function related to the all order PT-BFM   gluon propagator  which in Landau gauge reads
\be
G^{\mu\nu}=\left[-g^{\mu\nu}+\frac{k^{\mu}k^{\nu}}{k^2}\right] d(k^2)
\ee
and satisfies usual Lehman representation 
\be \label{spectral}
d(k^2)=\int_0^{\infty} d\om \frac{\rho(\om)}{k^2-\om+\ep}\, 
\ee
and $\tilde{\Gamma}_T$ is the rest of the three gluon unproper vertex which is not specified by gauge invariance.
In the expression (\ref{spectral}) we assume the branch point is located in the beginning of complex plane. Albeit not generated by poles, 
it  is in its usual ``non-anomalous ``  position.  

The essential feature of the vertex $\Gamma_{\nu\alpha\beta}$ is that apart from the structure dictated by WTI  it also includes $1/q^2$ pole term which gives rise to infrared finite solution. For this purpose the following form 
\bea \label{YMSCHWINGER}
d(k)\tilde{\Gamma}_T^{\nu\alpha\beta}(k,q)d(k+q)&=&\int d\om \rho(\om)\frac{1}{k^2-\om+\ep}\Gamma_{\nu\alpha\beta}^T
\frac{1}{(k+q)^2-\om+\ep} \,
\nn \\
\Gamma^{\nu\alpha\beta}_T(k,q)&=&c_1[(2k+q)_{\nu}+\frac{q_{\nu}}{q^2}(-2k.q-q^2)]g_{\alpha\beta}
+[c_3+\frac{c_2}{2q^2}((k+q)^2+k^2))](q_{\beta}g_{\nu\alpha}-q_{\alpha}g_{\nu\beta}) \, ,
\eea
has been proposed in \cite{AGPA2006}. This vertex is transverse with respect to $q$ ($q.\Gamma=0$) and it  respects Bose symmetry 
to two quantum legs interchange.

After the renormalization, it leads to the following form of  SDE in Euclidean space:
\bea \label{haf}
d_E^{-1}(q_E^2)=q_E^2\left\{K+bg^2\int_0^{q_E^2/4} dz \sqrt{1-\frac{4z}{q_E^2}} d_E(z)\right\}
\nn \\
+\gamma bg^2 \int_0^{q_E^2/4} dz z \sqrt{1-\frac{4z}{q_E^2}} d_E (z)+d_E^{-1}(0) \, ,
\eea
where the second line appears due to the Ansatz for the gluon vertex (\ref{YMSCHWINGER}) and $K$ is the renormalization constant.
Thus, the strength of the dynamical mass generation is triggered through  the  adopted coupling constants $c_1,c_2,c_3$,  
which is  fully equivalent to introducing (in principle arbitrary)  constant $\gamma$ 
and infrared value $d_E^{-1}(0) $.

Analytical continuation $q^2_E= -q^2=-s \rightarrow q^2 , s=q^2>0$ of the Eq. (\ref{haf}) is  very straightforward 
 and  the   gap equation for the gluon propagator for timelike variable $s$ reads
 \bea \label{haf2}
d^{-1}(s)=s\left\{K+bg^2\int_0^{s/4} dz \sqrt{1-\frac{4z}{s}} d(z)\right\}
+\gamma bg^2 \int_0^{s/4} dz z \sqrt{1-\frac{4z}{s}} d (z)+d^{-1}(0)+i\epsilon \, \, \, ,
\eea
 where we use standard convention $d_E(q_E^2)=-d(s) $ for $  s=q^2<0 $ and absorb the sign into the definition of Minkowski space
 gluon propagator.  Note, the inverse propagator $d^{-1}(0)$ 
 should take a negative value to prevent tachyonic solution.

 Subtracting the equation once again at the timelike fixed scale $\zeta$ one gets:
\bea \label{mnau}
d^{-1}(s)=s\left\{K+bg^2\int_0^{s/4} dz \sqrt{1-\frac{4z}{s}} d(z)\right\} -(s\rightarrow \zeta)
\nn \\
+\gamma bg^2 \int_0^{s/4} dz z \sqrt{1-\frac{4z}{s}} d (z) -(s\rightarrow \zeta) -d^{-1}(\zeta) \, \, \, .
\eea
Now, the function $d$ is assumed to be complex for all $s>0$, while it stays real for negative $s$. 

Like in previous study of strong coupling QED, the phase of the propagator needs 
to be tuned to get desired analytical property.
The numerical search of the phase of the function $d(\zeta)$ is performed such that we   fix the real part $Re d^{-1}(\zeta)$ and then  look for the value of  $\Im d^{-1}(\zeta)$ till assumed integral representation, e.g.
the following relation 
\bea \label{hurvajz}
\Re d(k^2)&=&\frac{-1}{\pi} P. \int d\om \frac{\Im d(\om)}{k^2-\om}\, ,
\eea 
 holds for all momenta.

\begin{figure} 
\centerline{\epsfig{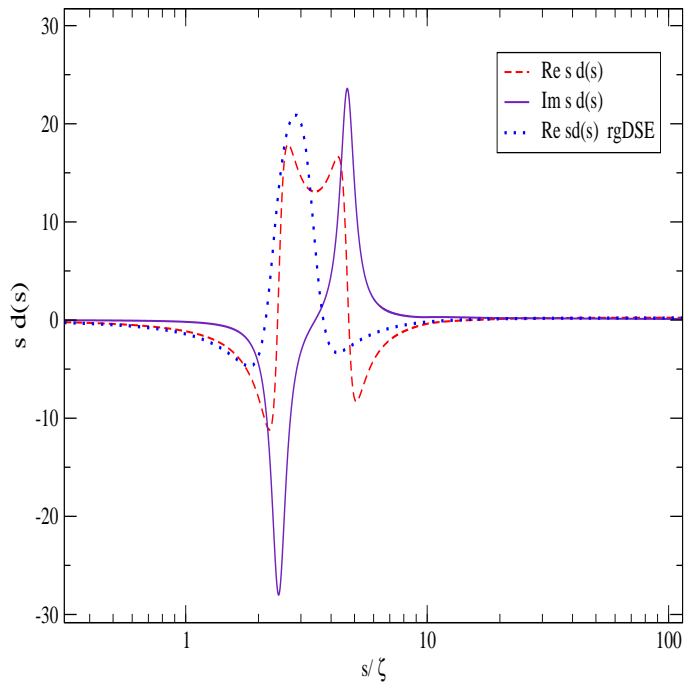}}
\caption[caption]{\label{gluon19}Gauge Technique governed SDE solution for gluon propagator plotted at the timelike  domain of momenta (in units of renormalization scale $\zeta$). The real part of the solution for the so called  Renormgroup improved SDE (see \cite{SAULI} for the meaning) is added for comparison. }
\end{figure}

Detailed  values of renormalization constant $K$ and $d^{-1}$ are not  crucial for the subject of confinement, 
however they need to be tuned to get agreement with the lattice data. As a first we  present the solution of SDE governed
by the Gauge Technique and take the transverse term very small  for this purpose, numerically we put $\gamma=1/20$
in the second term in (\ref{mnau}).  Further we set the coupling $g^2=15.5 $ 
and for renormalized propagator the condition  $\Re \zeta d(\zeta)=1.8570$ and $K(\zeta)=1$  was chosen.  
Searching for correct analyticity we obtained  $\Im  \zeta d(\zeta)=0.140$ at fixed $\zeta=1GeV$  after the iterations.
 This Gauge technique governed solution is shown in the Fig. \ref{gluon19} for the  timelike momenta.
Note the smooth solution for the spacelike momenta is not shown, and we just note it is  
  is several times smaller the lattice data in the infrared domain, hence non-compatible at all.

\begin{figure} 
\centerline{\epsfig{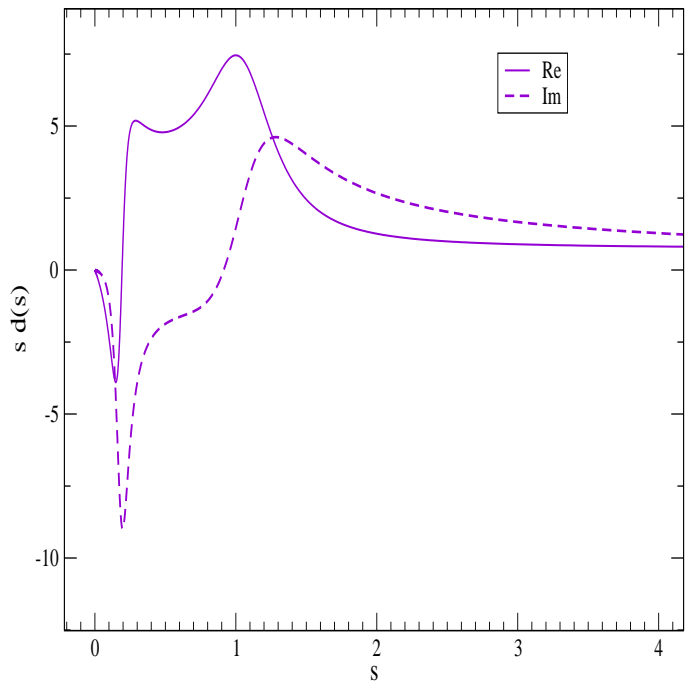}}
\caption[caption]{\label{gluon23} Lattice like  solution of SDE for the gluon propagator plotted at the timelike  domain of momenta against $s=p^2$ in units of $GeV^2$. The first peak is located at $445 MeV$. }
\end{figure}

\begin{figure} 
\centerline{\epsfig{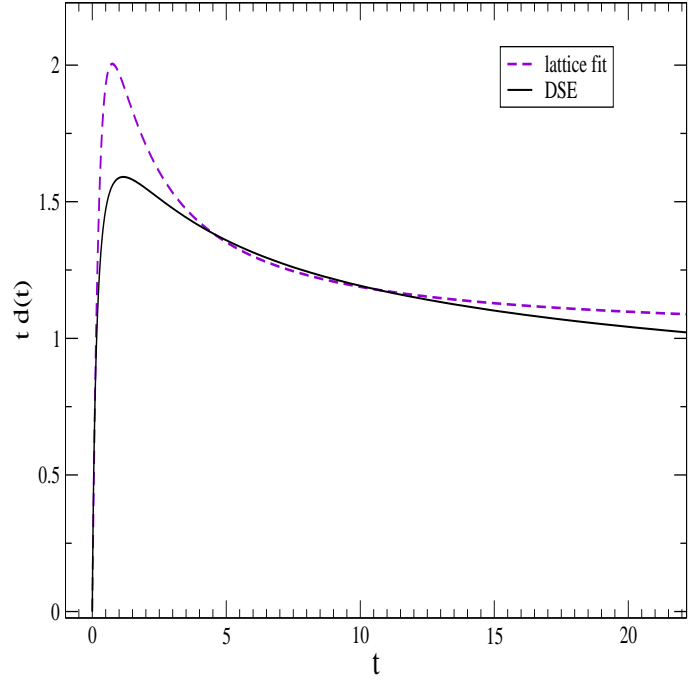}}
\caption[caption]{\label{gluon17} Lattice like solution of gluonic  SDE compared to the conventional lattice fit \cite{fitc,coim2}   in the Landau gauge. }
\end{figure}

 The solution, which complies reasonably with the lattice data  \cite{fitc,coim2,coim3} is shown in figures \ref{gluon23} and \ref{gluon17}
 where in the later case the comparison with lattice fit is  made.  A relatively large coupling for  transverse component of the 
 vertex was needed to make the solution comparable to lattice gluon propagator in the infrared.
 The excellent small error difference  $\sigma^2=3.5 10^{-5}$ was achieved 
 for difference between lhs and rhs of equation (\ref{hurvajz}), showing that desired analytical properties can be achieved with 
 high accuracy. Obviously the SDE solution and the lattice data doesn't agree completely as our transverse component was modeled
and not a complete vertex. Better agreement between the lattice and the SDE solution would require approximation improvement, e.g. by inclusion of not yet considered  transverse component of triple vertex component.  Perhaps  the inclusion of ghost loop should improve numerical agreement as well.

 Numerically, our    solution was obtained  within the following  couplings: $\gamma=-4.5$ and $g^2=2.167$, and 
 the renormalization and complex phase was determined by   values $\Re  d(\zeta)=8.323*10^{-2} GeV^2$ ; $\Im  \zeta d(\zeta)=7.58 10^{-3} GeV^2$
 at the timelike renormalization scale $\zeta=2.89 GeV$. In this case we found advantageous to reduce the constant, such that $\Re K=1/2$ was taken and 
 $\Im K= 7.3*10^{-2}$ was  iteratively searched value.
  Codes running at the end of iteration cycles are available at author's web page (questions can be  asked via email).

%$ with very similar, albeiti not identical program is  rez2$
   
At ultraviolet the studied gluon propagator vanishes as $1/s ln^{\kappa}(s)$ with $\kappa\simeq 0.5$ 
 (the size of  $\kappa$ is understand mainly  because of  Landau gauge used in primitive truncation here). This behavior should lead to 
 super-convergent relation sum rule \cite{OZ1980},\cite{OEM1990},\cite{COR2013}:
\be \label{nuf}
I=\int_0^{\infty} \rho(s) ds=0 \, ,
\ee 
which turns to be satisfied with reasonable accuracy $I\simeq 0.05$

In both presented numerical solutions: the Gauge Technique like and lattice like one, the shape of the gluon propagator  at the timelike region is obviously something what we are not experienced (see the Fig. \ref{gluon19} and the Fig. \ref{gluon23} respectively).
The real part of the function $d$ is represented by two broad peaks with mutually opposite signs.
However let us recall here, that  kind of the resonance  with zero mass anomalous threshold  of  the form
\be    \label{schw}
\rho(s)=\frac{ C s}{(s^2-s_o^2)^2+(s\Gamma)^2}
\ee
was suggested as an artificial mathematical model for the spectral function of the photon in the  Schwinger model \cite{SCHWING}.
Interestingly,  the spectral gluon  function can be accurately estimated as a difference of two such excitations.
Because of the asymptotic freedom, instead of using (\ref{schw}),
 more suited fit can be constructed from the sum of  modified Cauchy distributions,  instead of using (\ref{schw}) and we take
\be
\rho(s)=\Sigma_i [ R\frac{(s/\zeta)^{\lambda}}{(s-s_o)^2+(\Gamma)^2}]_i \, ,
\ee
 where the exponent $\lambda=0$ for the positive modes of the function $\rho$ (negative modes in  figures), while for 
 the  negative mode the exponent  ${\lambda=3/4}$ was taken. 
 Only two such contribution are  enough to describe the  gluon propagator  in the infrared domain of momenta.  Fitting the first peak structure  at 800MeV,
 the second  peak   then appears at 1110 MeV for Gauge Technique governed solution. 
 
 More interestingly, performing similar fit for the lattice like solution,  one gets the position of the first peak  $\mu_g=450 MeV$, while much broad peak
 of opposite sign is located nearly at $1 GeV$ (see again the Fig. \ref{gluon19}).   For the gluon mass scales estimated by others, see for instance \cite{COR1982} $\mu_g=500\pm 200 MeV$ and \cite{AFTP2019} $\mu_g\simeq 0.5 GeV$.

-
\\
-
\\

Let us stress here that the gluon propagator is intrinsically unphysical quantity and  neither of  fit presented here follows any experimental data.
 All presented scales (which is the only number wee need for a given fit) could be understand as educated estimate coming from gauge  invariant  considerations  -string tension, meson masses or  decay constants- performed in the Landau gauge. 

Actually, among other choices, it is  the Landau gauge that  was extremely popular and preferred
in SDEs studies during last decades. However, as very well known (see for instance \cite{alk2021})   changes within  varied gauge fixing parameter are shown to be dramatic in the sector of unphysical gluon and ghost propagators. Hence to understand confinement in QCD properly, one could be able to build 
gauge (fixing) invariant quantity in framework of  SDEs and to the best provide the hadronic form factor for arbitrary gauge.  To this point a  departure form popular Landau gauge turns to be an advantageous choice.  We briefly mention a limited  progress made in this direction in the following  section.

 \section{Short remark on the quark propagator}

Following similar method described in the Sect. 2, the ladder-rainbow approximation (LRA) for the quark SDE has been already
solved in the paper \cite{sau1} in other gauges then Landau gauge.
 The obtained quark propagator  has an anomalous threshold located at or very near beginning of 
complex plane of square of momenta and the confinement of quarks manifests as missing real pole in the quark propagator.
Interested reader can find the details in the paper  \cite{sau1}.

  As an interesting task  we have accomplished the same approximation as in the paper \cite{sau1} , but  with kernel 
given by the Landau gauge gluon propagator instead. As a matter of fact, such approximation neither provides a
 known slope of the  quark dynamical mass function or correct pion observable (mass and decay constant $f_{\pi}$). Hence neither 
 such approximation can be used to calculate continuous form factor. 
 A possible    understanding of this inefficiency  is obvious from the
  modern version of   Goldberger-Treiman-like  relation \cite{MAROTA1998},\cite{QIROS2014},
   which  reads  
  \be \label{GTR}
 f_{\pi} \Gamma_A(0,k)=B(k)
 \ee
  where $\Gamma_A$ is the leading piece of the pion’s Bethe-Salpeter amplitude and $B$ is the quark dynamical mass times
the quark renormalization function.  As can be seen form the quark DSE only trivial solution for  the rhs. of E. (\ref{GTR}) exists.
In other words, a  single gluon exchange approximation with Landau gauge gluon propagator underestimates of the true strength of the quark-antiquark interaction.

Miraculously, the picture is dramatically changed when one leave the Landau gauge and utilize exactly the same method but 
 for larger gauge fixing parameter. The pions are then true  Goldsteone bosons correctly  generated  in such model within LRA ,were the gluon propagator is strengthened by presence of large longitudinal modes in the quark-antiquark interacting kernel. 
 Actually a few existing  calculations provide first hints for working  scheme of SDEs solutions within a sort of integral representations for QCD/QED Green's functions. Of course, the calculation of continuous form factors are not easy in presented scheme and various approximations
 limit the method in practice at these days. The reader can find the first  applications in calculation of meson transition and the meson electromagnetic  
 form factor in \cite{sau2,sau3,YPNFS2021}

\section{Conclusion}

We have applied the method of subtractions to SDEs at the timelike scale  of momenta and  get the confined solution 
in Yang-Mills theory for the gluon propagator as well as for the fermion propagator in quenched QED. In both strong coupling models, the solutions for  two point correlator were obtained 
in the entire domain of Minkowski space momenta. 
When used in other gauges or in more sophisticated approximations,  the  method is a possible candidate for numerical evaluation of hadronic form factors needed to describe a  production processes in hadronic physics. 
We do not incorporate complex conjugated poles, which are nowadays seen numerically 
 in spectra of lattice propagators of confined fields \cite{BT2019}. They were not  necessary for solutions presented here.
On the other side, if complex conjugated poles are not a gauge dependent and numerical artifact and they 
will become  a rigid  analytical structure of QCD correlator, they need  to be  incorporated properly. While, this is certainly possible by deforming 
the integration contours, the presence of them requires access to entire complex plane and would make the evaluation extremely demanding.
For a  nice discussion  of this open problem, see  recent paper \cite{HPW2022},
where  consequences of embedding  the complex conjugated poles  into the gluon propagator are shown for a simple truncation of ghost/gluon DSEs system 
 in a pure  Yang -Mills  sector.  Obviously, a search of calculation scheme with limited amount of non-holomorphic singularities is worthwhile to search in QCD and other strong coupling quantum field models with confinement as well.

\appendix

\section{Appendix A}

The second constrains which should be checked beyond ladder-rainbow approximation can be 
derived by making the trace of $\not p$-projected fermion propagator, where 
\be \label{equiv2}
L_v(s)=R_v(s)=\frac{1}{4p^2} \Re \not p Tr S(s)=\Re S_v(s) \, .     
\ee
 with the left and right hand sides  defined as 
\bea \label{cond2}
L_v(s)&=&P.  \int_0^{\infty}d a\frac{\rho_v(a)}{s-a}
\nn \\
R_v(s)&=&\frac{-\Re A(s) R_D(s)+\Im A(s) I_D(s)}{R_D^2(s)+I_D^2(s)}         
\eea  
where symbol $P.$ stands for principal value integration.

\end{document}